\documentstyle[aps] {revtex}
\begin{document}
\title{Ewald methods for polarizable surfaces with
application to hydroxylation and hydrogen bonding on
the (012) and (001) surfaces of $\alpha$-Fe$_2$O$_3$}

\author {E.~Wasserman$^1$\footnote{Author 
to whom correspondence should be
addressed. E-Mail: e\_vasserman@ccmail.pnl.gov, fax (509) 372-6328, 
phone (509) 372-6083 }, 
J.~R.~Rustad$^1$, A.~R. Felmy$^1$,  B.~P. Hay$^1$ and  J.~W. Halley$^2$}
\address{
$^1$
Pacific Northwest National Laboratory  ETB-K9-77, 3200 Q Avenue \\
Richland, WA 99352 \\
$^2$Department of Physics and Minnesota Supercomputer Institute \\
University of Minnesota, Minneapolis, MN 55455}

\date{\today}
\maketitle

\begin{abstract}
We present a clear and rigorous derivation of the Ewald-like method
for calculation of the electrostatic energy of the systems infinitely
periodic in two-dimensions and of finite size in the third dimension
(slabs). We have generalized this method originally developed by Rhee
{\it et al.} [Phys. Rev. B 40, 36(1989)] to account for charge-dipole
and dipole-dipole interactions and therefore made it suitable for
treatment of polarizable systems. This method has the advantage over
exact methods of being significantly faster and therefore appropriate
for large-scale molecular dynamics simulations. It however involves a
Taylor expansion which has to be demonstrated to be of sufficient
order.  The method was extensively benchmarked against the exact
methods by Leckner and Parry. We found it necessary to increase the
order of the multipole expansion from 4 (as in original work by Rhee
{\it et al.})  to 6. In this case the method is adequate for 
aspect ratios (thickness/shortest side length of the unit cell) $\leq
0.5$.  Molecular dynamics simulations using the
transferable/polarizable model by Rustad {\it et al.}  were applied to
study the surface relaxation of the nonhydroxylated, hydroxylated,
and solvated surfaces of
$\alpha$-Fe$_2$O$_3$ (hematite).  We find that our nonhydroxylated
structures and energies are in good agreement with previous
LDA calculations on $\alpha$-alumina by Manassidis {\it et al.}
[Surf. Sci. Lett. 285, L517, 1993].  Using the results
of molecular dynamics simulations of solvated interfaces,
we define end-member hydroxylated-hydrated states for 
the surfaces which are used in energy minimization calculations.
We find that hydration has a small effect on the surface
structure, but that hydroxylation has a significant effect.
Our calculations, both for gas-phase and solution-phase adsorption,
predict a greater amount of hydroxylation for the $\alpha$-Fe$_2$O$_3$
(012) surface than for the (001) surface.  Our simulations
also indicate the presence of four-fold coordinated iron ions
on the (001) surface.

\end{abstract}

\section{Introduction}
In a wide variety of scientific and engineering problems, simulation
techniques connecting models at \AA ngstrom scales to `microstructural'
scales 10$^4$-10$^5$ times larger are required.  Classical simulation
of atomic and molecular systems works between these two extremes
and is useful for moving atomic scale predictions into the
microstructural or thermodynamic regime 
\cite{philpott-glosli-zhu95,rustad-hay95,rustad-felmy-hay96b}. 

Here, we consider simulations using 
models with analytical interaction potentials 
involving charges and dipoles in interfacial environments.
The long-range coulomb interactions in such systems must typically
be taken into account over distances much larger than the sizes 
of samples which can be directly simulated.  Historically this problem
has been approached by infinitely and periodically repeating the simulation
sample in all three dimensions.  The long-ranged contributions to the energies
and forces can be determined by a variety of methods.
They include transformation to rapidly convergent sums in real space
\cite{leckner91}, the fast multipole method 
\cite{greengard-rokhlin87} and various Ewald-type methods 
\cite{leckner91,parry75,heyes84,rhee-halley-hautman-rahman89,hautman-klein92}.
The latter involve summations over the reciprocal lattice of the 2-D
lattice. The relative merits of different methods are reviewed by
Tildesley \cite{tildesley93}.  The original 2-D Ewald method
due to Parry \cite{parry75} is commonly used in quantum
mechanical calculations where the Ewald summation comprises
a small fraction of the computation time.  For simulations
with  parameterized analytical potentials, using Parry's method
makes the Ewald summation procedure rate-limiting
and it becomes worthwhile to explore 
alternatives to Parry's method.

In the present paper we extend the 2-D periodic Ewald method of 
reference \cite{rhee-halley-hautman-rahman89} to include polarizable atomic 
centers.  
Effective pairwise potential energy functions parameterized for bulk systems
do not work well at surfaces because
the surface intermolecular environment is entirely different from
that of the bulk.  The first step in enhancing the
transferability of empirical potentials is to consider explicitly
the effects of electronic polarization \cite{barnes-finney-nicholas-quinn79}. 
Therefore, the extension to polarizable systems is quite important
in a method designed to be used at interfaces.  In addition, we have
fixed some omitted fourth order terms in reference \cite{rhee-halley-hautman-rahman89} and
we have carried
out the Taylor expansion to
sixth order in the aspect ratio, significantly improving the accuracy
of the method.
We have also clarified the extension to
charged systems.

The generalized method is
tested against previously-published
benchmarks, and then applied to investigate the structure
and hydroxylation of $\alpha$-Fe$_2$O$_3$ (hematite) surfaces. 
Hematite is chosen because it (as well as its aluminum analog, corundum) 
is a good example of a material which is well-studied using high-vacuum 
techniques and state of the art theoretical methods, and is also important
in natural weathering and adsorption 
processes of interest in soil chemistry 
\cite{nooney-murrell-corncille-wood-goodman95}.   

\section{Methods}

Our polarizable system is most conveniently treated by generalizing
the expressions for charge-charge interactions from
Ref.~\cite{rhee-halley-hautman-rahman89} to include charge-dipole and
dipole-dipole energies and forces.

We start by  deriving the Ewald-type expression for the energy of a system 
of point charges periodic in two dimensions and finite along the 
third dimension.   If the monopole-monopole potential energy 
expression is given by $\Phi_{mm}(r,s,z)$, then one can
work out the expressions for the charge-dipole, and
dipole-dipole interactions using:
\begin{equation}
\Phi_{md}(r,s,z)=-q_j\vec\mu_i\cdot\nabla\Phi_{mm}(r,s,z)
\label{phicd}
\end{equation}
for the charge-dipole interaction,
and
\begin{equation}
\Phi_{dd}(r,s,z)=-(\vec\mu_j\cdot\nabla)(\vec\mu_i\cdot\nabla)\Phi_{mm}(r,s,z)
\label{phidd}
\end{equation}
for the dipole-dipole interaction, as well as
gradients involving these terms arising in the
expressions for the forces \cite{deleeuw-perram-smith80a}.

The derivation of the Ewald summation for the case of two-dimensional slabs of
finite thickness is given in previous publications
\cite{rhee-halley-hautman-rahman89} and \cite{hautman-klein92}. We 
present another method of  derivation 
that we believe 
is easier to follow and rigorous. It is essentially a special case of the derivation
presented in
\cite{hautman-klein92}.

The electrostatic energy of the system of point charges with 2-D periodicity  is
given by 
\begin{equation}
 \Phi_{mm}=\frac{1}{2}\sum_{\vec \nu} { 
\sum_{i,j}
{ ^\prime \frac {q_i q_j}{|\vec r_{ij}+\vec \nu|}} 
-\frac{1}{2}(\sum_{i=1}^N {q_i})^2 \frac{1}{A} \int \frac{d\vec s}{s}  }
\label{eq:start}
\end{equation}
Here $\vec r_{ij}$ is the separation between particles $i$ and $j$,
$\vec s_{ij}$ is the projection of this vector on the plane, $\vec\nu$
is the translation vector in the plane.  $N$ is the total number of
particles, and $A$ is the surface of the 2-D unit cell.  The last term
is the energy of the interaction of the slab (if non-neutral) with the
uniform neutralizing background.  The prime on the summation sign
means that the term $i=j$ is skipped when $\vec \nu=0$. A vector
quantity without an arrow denotes the magnitude of the vector.

As in the 3-D Ewald method, we
split the infinite sum (\ref{eq:start}) into  short-ranged ($V_S$) and 
long-ranged  ($V_L$) parts. $V_L$ will be replaced by its Fourier series 
expansion.  To perform this separation, we write the identity 
\begin{equation}
\frac{1}{r}=\frac{1}{r}-\frac{h_0(s,\gamma)}{s}+\frac{h_0(s,\gamma)}{s},
\end{equation}
where $h_0(s,\gamma)$ is the 
convergence function with a parameter $\gamma$ (to be defined below). 
One then obtains:
\begin{equation}
\Phi_{mm} = V_S+V_L,
\end{equation}
\begin{equation}
V_S=\frac{1}{2}\sum_{\vec\nu} {\sum_{i,j}
{ ^\prime q_i q_j \left(1/r_{ij\nu}-h_0(s_{ij \vec \nu}, 
\gamma)/s_{ij\nu}\right) } }
\label{eq:short}
\end{equation}
\begin{equation}
V_L=\frac{1}{2}\sum_{\vec\nu} { \sum_{i,j}
{ ^\prime q_i q_j h_0(s_{ij\nu}, \gamma)/s_{ij\nu}} 
-\frac{1}{2}(\sum_{i=1}^N {q_i})^2 \frac{1}{A} \int \frac{d\vec s}{s}
}
\label{eq:long}
\end{equation}

The cutoff function $h(s)/s$ is defined by specifying its Fourier image. 
Consider the  function $w(s) \equiv \sum_{\vec \nu} h_0(s_\nu)/s_\nu$, where 
$\vec s_\nu=\vec s + \vec \nu$ and $s_\nu=|\vec s_\nu|$.
It is periodic with respect to 
the 2-D   translation vector $\vec \nu$. Therefore it can be expanded in 
Fourier series in  2-D reciprocal space vector $\vec G$, $C_{\vec G}$ 
being the Fourier coefficients: 
\begin{equation}
w(s) \equiv \sum_{\vec \nu}{ h_0(s_\nu)/s_\nu} 
=\sum_{\vec G}C_{\vec G} \exp(i\vec G \cdot \vec s)
\label{eq:expansion}
\end{equation}
The sum is over all reciprocal lattice vectors, including $\vec G=0$. 
Following  the same notation as in Ref.~\onlinecite{hautman-klein92}, 
we define the function $f(G)$ from 
\begin{equation}
C_{\vec G} = \frac{2 \pi}{A} \frac {f(G)}{G},
\label{coefs}
\end{equation}
where $A$ is the area of the 2-D unit cell. 
The Fourier coefficients $C_{\vec G}$ of the 
function $\sum_{\vec \nu} h_0(s_\nu)/s_\nu$ are
\begin{equation}
C_{\vec G} = \frac{1}{A} \int_A{
\sum_\nu{\frac{h_0(s_{\nu})}{s_\nu}}
\exp(-i\vec G \cdot \vec s) d \vec s
                               }.
\label{eq:series}   
\end{equation}
Changing the order of summation and integration and using the additivity of 
integration, one can write
\begin{equation}
C_{\vec G} = \frac{1}{A}\int_{\rm all\ s-space}
{ \frac{h_0(s)}{s} \exp(-i \vec G \cdot \vec s) d \vec s}
\end{equation}
Comparing the expression above to 
the definition of the 2-D Fourier transformation  
of the function  $h_0(s)/s$,
\[
\frac{1}{2 \pi} \int_{\rm all\ s-space} {\frac{h_0(s)}{s} 
\exp(-i\vec G \cdot \vec s) d \vec s}, 
\]
one obtains
\begin{equation}
\frac{f(G)}{G}=\frac{1}{2\pi} \int_{\rm all\ s-space} {\frac{h_0(s)}{s}}
\exp(-i\vec G \cdot \vec s) d \vec s.
\label{eq:fourier}
\end{equation}
This means that $f(G)/G$ is the 2-D Fourier image of  $h_0(s)/s$. 
Following Ref.~\onlinecite{rhee-halley-hautman-rahman89}, we choose 
\begin{equation}
f(G)=\exp[-(\gamma G)^3].
\end{equation} 
The direct space cutoff function $h_0(s)/s$ is obtained by 
transforming back from $\vec G$-space to $\vec s$-space
\begin{equation}
\frac{h_0(s)}{s}=\frac{1}{2 \pi} \int_{\rm all\ G-space}
{\exp(i\vec G \cdot \vec s)}
\frac{\exp[(-\gamma G)^3]}{ G} d {\vec G}
\end{equation} 
This integration is conveniently done using 
 polar coordinates in the $\vec G$-plane:
\begin{equation}
 \frac{h_0(s)}{s}=\int_0^{\infty}G d G
\left[\frac{1}{2 \pi}\int_0^{2 \pi}{\exp(iGs \cos\theta)} d \theta \right]
{\frac{\exp[(-\gamma G)^3]}{G}} 
=\int_0^{\infty}J_0(G s) \exp(-\gamma^3 G^3) dG.
\label{eq:image}
\end{equation}
Here we used the well-known integral
\begin{equation}
\frac{1}{\pi} \int_0^{\pi} {\exp(ix \cos \theta) d \theta}=J_0(x),
\end{equation}
where $J_0(x)$ is the Bessel function. 
$V_L$ will be calculated as the sum of Fourier series. To do so,
rewrite it using the periodic function $w(s)$ defined above. 
This is done by adding and subtracting $i=j$ term. 
\begin{equation}
V_L=\frac{1}{2} \sum_i {\sum_j{q_i q_j w(s_{ij})}}
-\frac{1}{2} \left( \lim_{s \rightarrow 0}
 {\frac{h_0(s)}{s}} \right) \sum_{i=1}^{N} {q_i^2}-
\frac{1}{2A} \left( \sum_{i=1}^{N} {q_i} \right) ^2 \int_{\rm all \ s-space} \frac{d\vec s}{s} 
\label{eq:long2}
\end{equation}
The limit in the expression above exists, and the direct calculation yields
\footnote{ \protect$\Gamma \left( \frac{1}{3} \right) = 2.67893853471$}
\begin{equation}
\lim_{s \rightarrow 0} \frac{h_0(s)}{s}=\frac{1}{3 \gamma}
\Gamma \left(\frac{1}{3}\right).
\label{eq:limit} 
\end{equation}
Substituting (\ref{eq:limit}) and (\ref{eq:image}) in 
(\ref{eq:long2}), one obtains
\begin{equation}
V_L=\frac{\pi}{A} \sum_{\rm all\ \vec G} { \frac{\exp(-\gamma^3 G^3)}{G} }
|\sum_{i=1}^{N}{q_i \exp(i \vec G \cdot \vec s)}|^2
-\frac{1}{6 \gamma}\Gamma\left(\frac{1}{3}\right)\sum_{i=1}^{N}{q_i^2}
-\frac{1}{2A} \left( \sum_{i=1}^N {q_i} \right)^2 
\int_{\rm all \ s-space} \frac{d\vec s}{s}
\end{equation}
This expression contains two infinite terms, 
the $\vec G=0$ term in the Fourier 
sum and the last term. The sum of these terms, 
$I=\int_{\rm all\ s-space}{\frac{h_0(s)-1}{s}d \vec s}$, can be written as 
\begin{equation}
I=\lim_{ G \rightarrow 0} { \left[
\frac{1}{A} 
\int_{\rm all\ s-space}{
\frac{h_0(s)}{s} 
\exp(-i\vec G\cdot \vec s) d \vec s }
-\frac{1}{A} \int_{all\ s-space}{\exp(-i\vec G\cdot \vec s) \frac{d \vec s}{s}}
   \right] }
\label{eq:depol}
\end{equation}
As one can see from the Eqs.~(\ref{eq:expansion}) and  (~\ref{coefs}), 
the first integral in (\ref{eq:depol})  is 
$\frac{2 \pi}{A}\frac{f(G)}{G}$. The second integral is calculated by
transforming to polar coordinates and using the formula
\[ \int_0^{\infty}J_0(bx) dx=1/b. \]
Its value is $\frac{2 \pi}{AG}$. 
Combining these expressions,  one obtains:
\begin{equation}
I=\lim_{G\rightarrow 0}{\frac{2 \pi}{A}
\frac{(f(G)-1)}{G}}=
\lim_{G\rightarrow 0}{\frac{2 \pi}{A}
\frac{(\exp(-\gamma^3 G^3)-1)}{G}}=0.
\end{equation}
Thus for our choice of $f(G)=\exp[{(-\gamma G)}^3]$, as well as other
functions $f(G)$ satisfying the above condition, the term due to the
background charge cancels exactly the $\vec G=0$ term of the
reciprocal space sum.  Therefore the long-ranged part of the
interactions can be written as
\begin{equation}
V_L=\frac{\pi}{A} \sum_{\vec G \ne 0} { 
\frac{\exp(-\gamma^3 G^3)}{G} 
\vert \sum_{i=1}^{N}{q_i \exp(i \vec G \cdot \vec s_i) \vert^2} }
-\frac{1}{6 \gamma}\Gamma\left(\frac{1}{3}\right)\sum_{i=1}^{N}{q_i^2}
\label{vl}
\end{equation}
Here we used the identity
\begin{equation}
\sum_{i,j} {q_i q_j \exp(i\vec G \cdot \vec s_{ij})}
=\vert \sum_{i=1}^N {q_i \exp(i \vec G \cdot \vec s_{i})}\vert^2 
\end{equation}

In practical calculations, the summations in real space are truncated at  
$|\vec \nu| = \vec \nu_{\rm max}$. The cutoff function $h(s)/s$ does not 
dissipate at large distances fast enough like it does for the 3-D Ewald 
method. Instead it has the asymptotic behavior $h(s)/s\approx 1/s$ at 
large distances.  
It is also convenient to separate the $i=j$ term in real space. 
We therefore write
\begin{equation}
\Phi_{mm}(r,s,z)=V_s^{\prime}+V_L+V_m
\end{equation}
\begin{equation}
V_s^{\prime}={{1}\over{2}}\sum_{i,j,j\ne i}
q_iq_j\sum_{\vec\nu=0}^{\nu \leq \nu_{\rm max}}
\left( {{1}\over{| \vec r_{ij}+\vec\nu | }}-
h(s_{ij \vec \nu, \gamma})/s_{ij \vec \nu}\right)
+{{1}\over{2}}\sum_i q_i^2\sum_{\vec\nu>0}\left(
{{1}\over{\nu}}-\frac{h_0(\nu)}{\nu} \right)
\label{vs}
\end{equation}
\begin{equation}
V_m={{1}\over{2}}\sum_{i,j,j\ne
i}q_i q_j\sum_{\nu > \nu_{\rm max}} 
\left({{1}\over{\vert\vec r_{ij}+\vec\nu\vert}}-{{1}\over{\vert\vec
s_{ij}+\vec\nu\vert}}\right)
\label{vm}
\end{equation}
The expression for $V_L$ is given by Eq.~(\ref{vl}).

The series for $V_m$ (Eq.~\ref{vm}) converges slowly, and direct 
calculation is extremely inefficient. Therefore it is advantageous 
to be able to precompute certain terms.
Defining the function $v_m^{\prime}$ 
\begin{equation}
v_m^{\prime}(r, s) = \sum_{\nu > \nu_{\rm max}} 
{\left({{1}\over{\vert\vec r+\vec\nu\vert}}
-{{1}\over{\vert\vec s+\vec\nu\vert}}\right) }, 
\end{equation}   
one can write 
\begin{equation}
V_m = \frac{1}{2} \sum_{i,j,i\ne j} v_m^{\prime}(r_{ij}, s_{ij}) 
\end{equation}
The key to the method of reference \cite{rhee-halley-hautman-rahman89}
is the use of a multipole expansion to
compute $V_m$.  The Taylor expansion of ${{1}\over{(1+x)^{1/2}}}$ is
\begin{equation}
{{1}\over{(1+x)^{1/2}}}=1-{{x}\over{2}}+{{3}\over{8}}x^2-
{{15}\over{48}}x^3+{{105}\over{384}}x^4+\ldots
\end{equation}
This, in conjunction with the elementary definitions of r and s:

\begin{eqnarray}
r & = & (\vert\vec s+\vec\nu\vert^2+z^2)^{1/2}\nonumber \\
& = &\nu\left({{s^2}\over{\nu^2}}+{{z^2}\over{\nu^2}}+2{{\vec
s\cdot\vec\nu}\over{\nu^2}}
+1\right)^{1/2}
\end{eqnarray}

\begin{eqnarray}
s & = & (\vert\vec s+\vec\nu\vert^2)^{1/2} \nonumber \\
&= &\nu\left({{s^2}\over{\nu^2}}+2{{\vec
s\cdot\vec\nu}\over{\nu^2}}+1\right)^{1/2}
\end{eqnarray}

and noting that any odd powers of $\vec s\cdot\vec\nu$ will vanish in the
sum on $\nu$, yields the following expression for the multipole
expansion in powers of (s,z)/$\nu$:

\begin{eqnarray}
v_m^{\prime} & = & -{{1}\over{2}}z^2\sum_{\nu=\nu_{\rm
max}}^\infty{{1}\over{\nu^3}}
+{{3}\over{4}}z^2s^2\sum_{\nu=\nu_{\rm max}}^\infty{{1}\over{\nu^5}}
+{{3}\over{8}}z^4\sum_{\nu=\nu_{\rm max}}^\infty{{1}\over{\nu^5}}
-{{15}\over{4}}z^2\sum_{\nu=\nu_{\rm max}}^\infty {{{(\vec s \cdot
\vec\nu)}^2}\over{\nu^7}}\nonumber \\
&&-{{15}\over{16}}(z^4 s^2+z^2 s^4)\sum_{\nu=\nu_{\rm max}}^\infty
{{1}\over{\nu^7}}
 -{{15}\over{48}}z^6\sum_{\nu=\nu_{\rm max}}^\infty {{1}\over{\nu^7}}
+{{105}\over{8}}z^2s^2\sum_{\nu=\nu_{\rm max}}^\infty{{{(\vec
s\cdot\vec\nu)}^2}\over{\nu^9}}\nonumber\\ 
&&+{{105}\over{16}}z^4\sum_{\nu=\nu_{\rm max}}^\infty {{{(\vec
s\cdot\vec\nu)}^2}\over{\nu^9}} 
\label{vmexpansion}
\end{eqnarray}
where we retain only the terms $s^nz^m$ for which n+m $\leq$ 6.

Application of equations~\ref{phicd} and~\ref{phidd} to obtain $\Phi_{md}$ and
$\Phi_{dd}$ is straightforward for $V_s$ (equation~\ref{vs}) 
and $V_L$ (equation~\ref{vl}), following
closely the 3-D periodic methods \cite{deleeuw-perram-smith80a}.
Reference~\cite{halley-rustad-rahman93}
gives a generalization of the method to polarizable systems.

To implement these equations
we split the interval of interest in 3 parts
and approximated $h_0(s)/s$ using the ratio 
of two polynomials.  The 3
derivatives of $h_0(s)/s$ were approximated separately, also
using rational approximation. The optimal rational approximation can
be easily constructed using the code given in Numerical Recipes
\cite{numerical_recipes}.

We report here the derivatives of $V_m$ (equation~\ref{vmexpansion})
in the Appendix, as these terms, while straightforward, are somewhat
tedious to compute.  We also note, as pointed out in reference
\cite{hautman-klein92}, that the expressions given in
\cite{rhee-halley-hautman-rahman89} do not contain all the terms to
the fourth order.  The expressions in the Appendix include multipole
contributions up to the sixth order.
Also, the numerical value of $\sum_{{\rm all}\ \vec \nu} {1/\nu^3}$
for the  square lattice with parameter $L$ as given  in reference
\cite{rhee-halley-hautman-rahman89} has only 3 correct digits and
should be replaced with $9.0333388/L^3$. This result was calculated
with quadruple precision. 

\section{Benchmarks and convergence test}
As one can see from the previous section, the Ewald-type method for
2-D slabs has 4 computational parameters: the convergence parameter
$\gamma$, cut-off parameters for summations in real and reciprocal
space and the order of Taylor expansion used to calculate $V_m$. In
this study we used $\gamma=0.1 L_{short}$, where $L_{short}$ is the
shortest side of the rectangular unit cell.  The summations in real
space included the first shell of neighboring cells, the total number
of cells included being 9.  
We increased the order of expansion from 4
(as was originally used by Rhee {\it et al}
\cite{rhee-halley-hautman-rahman89} and also used in
Ref.~\cite{hautman-klein92}) to 6, and this improved the
accuracy of the calculations.

A valid test of convergence with respect to all computational
parameters is a benchmark against another method. A test example for
the monopole-monopole interaction was provided in
Ref.~\cite{tildesley93}. As discussed in that
review, the method of Leckner
\cite{leckner91} has only one computational parameter, the cutoff of the
 summation in real space. It has exponential convergence and can be 
made as exact as required by choosing this cutoff appropriately.
Therefore it is ideal as a benchmark \cite{tildesley93}. The
comparison of the forces  and potential energies
calculated in this study with  those from Ref.~\cite{tildesley93} and
calculated using the implementation of Parry's method \cite{parry75} is
given in Tables~\ref{bench1} through \ref{bench3}. 
As one can see from Tables~\ref{bench1} and \ref{bench2},  
the two benchmark methods (Leckner's and Parry's) 
yield exactly the same results. One can therefore be confident 
in the benchmark itself. 
The results of our method are in excellent agreement with the benchmark 
for small interparticle separations in $z$ direction. Once the ratio of the 
maximum separation in $z$ direction to the magnitude of the translation
 vector (aspect ratio of the slab) increases, the agreement degrades 
slightly, the maximum error being 2.5 \% for the aspect ratio 0.5. 
The Taylor expansion used here to efficiently 
evaluate  $V_m$ is in powers of $(z_{ij}/\nu)^2$.  
Therefore the quality of representation of $V_m$ degrades as the 
expansion parameter increases. 
 This explains the results of the benchmark. The error of 2.5 \% in $F_z$ 
can be considered acceptable. Real molecular dynamics simulations 
are done with $\sim$1000 atoms, not with 2 atoms.  As noted 
by Tildesley \cite{tildesley93}, the convergence of 2-D 
Ewald-type methods improves when the cell is filled with charges, so that
ions on opposite faces of the slab are shielded from each other. 

We performed a test on the slab of 216 formula units of Fe$_2$O$_3$
(1080 atoms) that corresponds to the neutral termination of the (012)
face. The electrostatic part of the total energy and forces in this
system were calculated using Parry's method and the method described
in this study. The calculations yielded the total energy of -2584.148
$e^2/$\AA\ and -2583.626 $e^2/$\AA\, respectively. The maximum
deviation in the magnitude of any force on any atom was 1.17 \%
of the exact value.
  
  To the best of our knowledge, the benchmarks for charge-dipole and
dipole-dipole interactions in 2-D periodic boundary conditions have
not been published. We therefore present them in Tables~\ref{bench4}
and \ref{bench_dd}, respectively. 
Once dipoles are present, the energies calculated using the Parry's method
\cite{parry75} and our method differ by an additive  constant. This stems from 
the fact that the surface terms are treated differently. The agreement
of forces is good. Since evaluation of charge-dipole and dipole-dipole
interactions involves successive differentiation of $V_m$, the order
of expansion decreases. Therefore it is not surprising that the
quality of calculated charge-dipole and dipole-dipole forces degrades
faster with the aspect ratio of the slab than does the force for the case of
charge-charge interactions.  Overall, the agreement is still totally
satisfactory.  Since the agreement with an independent method was
good, we conclude that the choice of the computational parameters is
adequate.

\section{Application to Hematite Surfaces: Background and Previous Work}

The morphology of hematite 
($\alpha$-Fe$_2$O$_3$) crystals is complex
\cite{barron-herruzo-torrent88,colombo-barron-torrent94,rochester-topham79b,reeves-mann91,schwertmann87,schwertmann-cornell91}, both in natural and
synthetic samples.  
The wide variety of crystal faces present
has tended to discourage atomistic interpretation of binding at
hematite surfaces, with some notable exceptions
\cite{barron-herruzo-torrent88,colombo-barron-torrent94}.  
This contrasts with goethite ($\alpha$-FeOOH), where there
have been several experiments and much discussion concerning  the role of
contiguous surface 
(monocoordinated) hydroxo groups in bidentate binding of oxyanions
\cite{rochester-topham79b,parfitt-russell-farmer75}.  

The properties of both hematite ($\alpha$-Fe$_2$O$_3$) and
corundum ($\alpha$-Al$_2$O$_3$)  surfaces {\it in
vacuo} have been studied  experimentally
\cite{kurtz-henrich83,reeves-mann91,henrich-cox94,gautiersoyer-pollak-henriot-guittet96,eggleston-hochella92} and by computer 
simulations \cite{pisani-causa-dovesi-roetti87,mackrodt-et-al87,causa-dovesi-pisani-roetti89,guo-ellis-lam92,manassidis-devita-gillan93,mackrodt92,godin-lafemina94}. 
In this section we consider both ideal vacuum-terminated and hydroxylated
(001) and (012) surfaces of hematite.  
These are by far the best studied, and among the most prevalent
surfaces for both hematite and corundum \cite{henrich-cox94,zoltai-stout84}.

As is the case for any oxide surface,
neither the extent of hydroxylation of hematite, nor
its  hydroxylated structure is well-understood \cite{brown90}.
Surface hydroxylation has been demonstrated
to occur on hematite by vibrational spectroscopy
\cite{rochester-topham79b,busca-lorenzelli80}, 
but the spectra are very complicated and
not easily interpreted.
Here, we offer an application of our 2-D
periodic Ewald summation method for polarizable systems to investigate
the surface structures using our previously-developed methods for
treating Fe-OH and Fe-OH$_2$ interactions
\cite{rustad-felmy-hay96b,rustad-hay-halley95}.

\section{Application to Hematite Surfaces: Simulation procedure}
The atomic interactions were described using parameterized
potential functions developed previously \cite{rustad-felmy-hay96a}.
Briefly, this model includes formal charges on all ionic species,
polarizable oxygen anions with a cut-off function to reduce induced
polarization at short interionic separations, and various two- and
three-body short-ranged terms. The functional form of all interactions
and the list of parameters is given in
Ref.~\cite{rustad-felmy-hay96a}. The lattice parameters of hematite at
zero pressure were determined in Ref.~\cite{rustad-felmy-hay96a} by
minimizing the total energy of the system with respect to the
magnitudes of lattice vectors and internal structural parameters. They
are listed in Table 3 of Ref.~\cite{rustad-felmy-hay96a} along with
experimental values. It is worth pointing out  that the O-O and H-O 
interaction parameters used here were  exactly the same 
as in the model for liquid
water and gas phase clusters of water molecules. Considering this
stringent transferability requirement the agreement of the calculated
lattice constants with experimental values can be considered good.
However,   the Fe-O bonds
predicted by the model are systematically about 3\%-4\% too long.

The simulations were done in two
different geometries: in ``sandwich'' geometry with 3-D periodic
boundary conditions (PBC) and in slab geometry. In simulations with
3-D PBC the length of the translational vector along $z$ direction was
3 nm for all faces.
Therefore the slabs were separated from each other by empty space in
$z$-direction ($\sim$1.5 nm thick).  

The atomic positions were relaxed 
by the following procedure:  The sample was heated to 400 K
and cooled to 0 K in 3-D molecular dynamics (MD) simulations 2 ps
long. Then the energy of the system was minimized with respect to
atom positions
using the individual damping method by Bennett \cite{bennett75}. In
this method, the equations of motion are solved as in standard MD but
the velocity $\vec{v}_i$ of an atom is set to zero if $\vec{F}_i \cdot
\vec{v}_i < 0$, $\vec{F}_i$ being the force acting on the
atom. This procedure is carried out until convergence of the
potential energy is reached. The 3-D minimal structure was used for the
initial configuration for 2-D energy minimization (with no periodicity
along $z$-direction). 
 
Our surface energies are calculated
using 
\begin{equation}
\gamma={{1}\over{2}}(E_{slab}-E_{bulk})/A
\end{equation}
where $E_{slab}$ is the potential energy of the
slab system, $E_{bulk}$ is the potential energy of
an equivalent number of Fe$_2$O$_3$ molecules in
the bulk, and $A$ is the surface area.

\section{Results}

\subsection{Nonhydroxylated Surfaces}

The main utility of our potential set is our ability to treat
surface hydroxylation.  However, because there is little basis
for comparison with other theoretical or experimental results on
hydroxylated surfaces,
we are compelled to test the performance of our potential set
on non-hydroxylated surfaces.  Even for these, there is little
information available on hematite specifically.  However, as mentioned
above, there has been a substantial amount of work on corundum
surfaces to which we can qualitatively
compare our results 
\cite{causa-dovesi-pisani-roetti89,guo-ellis-lam92,manassidis-devita-gillan93,godin-lafemina94,mackrodt92}.  
We limit the investigation to the structure and
energetics of the (001) and (012) surfaces, as
these are the surfaces receive the most attention
both theoretically and experimentally.

\subsubsection{(012) Surface}

The (012) surface
has been previously studied in references
\cite{henrich-cox94,mackrodt-davey-black-docherty87,gautiersoyer-pollak-henriot-guittet96,guo-ellis-lam92}.
The (012) surface is often indexed using
the symmetrically equivalent 
${\rm (1\bar 102)}$ surface in hexagonal notation;
also note that in reference \cite{henrich-cox94}
this surface is indexed as ${\rm (10\bar 12)}$.  A cross
section of the (012) slab, as viewed down [100] is shown in 
Figure~\ref{012.nohydrox}(a).
This surface can be built up from neutral O-Fe-O-Fe-O units as described
in reference \cite{guo-ellis-lam92}.  The slab in Figure~\ref{012.nohydrox}(a)
is comprised of three such stacking units.  The unit cell predicted
 by the model is
0.556 nm $\times$ 0.518 nm.  For the calculations here, we use a supercell
consisting of 6 $\times$ 6 unit cells; the supercell is periodic in X and Y directions.

The atomic coordinates for the unit cell of the slab are given in
Table~\ref{012vacuum}. In Figure~\ref{012.nohydrox} we
show the two side views and the map view of the relaxed (012) 
top stacking unit (including the top O layer from the middle stacking unit).
The noteworthy zig-zag oxygen ridges characterizing
the (012) surface are clearly seen in Figure~\ref{012.nohydrox}(b).
The ``slopes'' of the oxygen ridges are occupied by five-fold coordinated iron
ions.  These are the darkly shaded iron ions in Figure~\ref{012.nohydrox}.
To test convergence of slab energies
and structures with respect to slab thickness, we considered systems having
one, two, three, and four O-Fe-O-Fe-O layers.  From Tables~\ref{012relaxstr} and
~\ref{gamma012}, one concludes that 3 layers is sufficient for convergence for both
energy and structure.
The thickness of the nonhydroxylated 3 layer slab, shown in Figure~\ref{012.nohydrox}
is 0.964 nm, and, for the size of the computational supercell considered here, the system
consists of 216 Fe$_2$O$_3$ molecules (1,080 total atoms).
The relaxation of each layer in the direction normal to the surface
is shown in Table~\ref{012relaxstr}.  Upon relaxation, the top
three layers are brought closer together, compensated by separation
of the bottom O-Fe layer.  The Fe-O distance across this
layer increases from 216 pm to 228 pm.

There is, to our
knowledge, only one theoretical study of $\alpha$ alumina (012)
\cite{guo-ellis-lam92}.  That work employed embedded-cluster
LDA methods to calculate energies of unrelaxed surfaces.  Our size-converged
unrelaxed surface energy of 2.8 J/m$^2$ (see Table~\ref{gamma012}) 
compares well with the value
of 2.95 J/m$^2$ obtained there.  
We find that allowing for relaxation 
decreases the surface energy to 2.0 J/m$^2$.

\subsubsection{ (001) Surface}

The (001) surfaces of both hematite and corundum have been
studied extensively \cite{henrich-cox94,gibson-lafemina96}.  We consider
here the standard neutral, nonpolar, nonhydroxylated termination based on the
stacking of Fe-O-Fe building blocks, as described in \cite{guo-ellis-lam92}.  
A cross section of the four Fe-O-Fe layer (001) slab, viewed down [100], 
is shown in Figure~\ref{001.nohydrox}(a).
The primitive hexagonal unit cell has a=0.518 nm; here we work
with a nonprimitive orthorhombic cell with dimension a=0.898 nm
and b=0.518 nm.  The computational supercell is 4$\times$6 
orthorhombic unit cells.  An orthorhombic cell was chosen for
computational efficiency.

The atomic coordinates for the unit cell of the slab are given in
Table~\ref{001vacuum}. In Figure~\ref{001.nohydrox}, we show the two
side views and the map 
view of the relaxed (001) surface.  We include the top two Fe-O-Fe
stacking units, but omit the bottom Fe layer for clarity.  In other
words, we show the top octahedral layer plus the overlying layer of
iron ions.
The model surface structure is characterized by a top layer three-fold coordinated
iron ions (darkly shaded in Figure~\ref{001.nohydrox}) in hexagonal
arrangement.
Configurations with two, three,
four, five, and six such blocks were used to test for convergence of
surface energy and surface structure as a function of the thickness
of the slab.  
Table ~\ref{001relaxstr} shows the percent change in distances
between layers parallel to the surface with respect to bulk distances.
Relaxation brings both the top Fe and O layers and the two interior Fe layers
closer together.

As shown in Table~\ref{quantum001_e}, our calculations are in
good agreement with previous calculations on isostructural $\alpha$ alumina,
\cite{mackrodt-davey-black-docherty87,manassidis-devita-gillan93,godin-lafemina94}
except that in the tight-binding calculations of reference
\cite{godin-lafemina94}, 
the distance between the two intra-octahedral
layers of Al ions increased by 20\%; in contrast to the
40-50 \%  decrease predicted by other empirical models and
LDA \cite{manassidis-devita-gillan93}.  The relaxed and unrelaxed surface
energies are reported in Table~\ref{gamma001}.  These compare
favorably with the energies for $\alpha$ alumina reported in reference
\cite{manassidis-devita-gillan93}, as well as the unrelaxed surface
energy reported in \cite{guo-ellis-lam92} (see Table~\ref{quantum001_z}).  
The structure, energy, and relaxation
energy of the (001) slab appear to be well-converged by the fourth, (or even
third) Fe-O-Fe layer.  The thickness of the four-layer slab, shown in Figure~\ref{001.nohydrox}(a)
is 0.773 nm. For the X and Y dimensions of the computational supercell,
the four-layer slab consists of 224 molecules (1120 total atoms).
The convergence with thickness is again similar to what was found for
$\alpha$ alumina in reference \cite{manassidis-devita-gillan93}.

Experimental fracture surface energies, as discussed 
in reference \cite{guo-ellis-lam92}, range between 24 J/m$^2$
and 6 J/m$^2$ for (012) to around 40 J/m$^2$ for (001).
The reason for the uniformly low values estimated by
all theoretical methods cannot be speculated on
at this time, but is a subject for future work.

\subsubsection{2-D Periodic vs. 3-D Periodic Calculations}

As one can see from Tables~\ref{gamma012} and~\ref{gamma001}, 
the minimal energy for the
case of 2-D PBC was always slightly higher than that for 3-D PBC.
This difference can however be considered small. We have therefore
demonstrated convergence with respect to the thickness of the empty
space for 3-D calculations. The difference in atomic positions between
3-D and 2-D minimal energy structures was no
more than 0.002 nm for all surfaces considered. The fact the 2-D
calculations agree well with the  3-D (which don't have any Taylor
expansion) in case of sufficient separations between slabs in $z$
direction suggests that for the actual system of interest the
convergence with respect to computational parameters was reached. 
This is of further interest because 3-D PBC need to be implemented in 
electronic structure methods incorporating plane waves 
\cite{manassidis-devita-gillan93}.  Our calculations support 
the inherent assumption that repeating the slab in the z direction
with sufficient empty space between the layers
has only a small effect on surface structures and energies 
for the systems considered here.  This conclusion
is of course system-dependent    
and would need to be demonstrated in any given case.

\subsection{Hydroxylated Surfaces}

In the previous section, it was demonstrated that our interaction
potential was in good agreement with other theoretical studies
for predictions on nonhydroxylated surfaces.  Extension to 
hydroxylated systems is quite difficult for both the
Hartee-Fock and LDA methods.  It has been demonstrated that 
treatment of hydrogen bonds 
requires very large basis sets and correlated wave functions 
for Hartree-Fock-based methods \cite{xantheas94} and gradient
corrections are required for density-functional methods \cite{laasonen-parrinello-car-lee-vanderbilt93}.
Our potentials are competitive with these methods. For example,
we obtain the same ordering of conformers of H$_3$O(H$_2$O)$_4$$^+$
as \cite{corongiu-kelterbaum-kochanski95} 
and the same ordering of the OH(H$_2$O)$_3$- as
\cite{xantheas95}.  Our model will provide
a semi-quantitative representation of surface
hydrogen bonding.

\subsubsection{Generation of the Hydroxylated Surface: Molecular Dynamics Calculations}

Defining the hydroxylated surface structure is in itself a difficult problem
\cite{brown90,johnsson-eggleston-hochella91}.
The coordination numbers of the iron atoms and extent of hydroxylation
are unknown. 
Because our potential model is as general as possible, we generate the 
hydroxylated surface structures through molecular simulation.  To begin
the simulations of the interfacial systems, we take the
vacuum terminations of the 216 formula unit (012) system 
(3 layers of O-Fe-O-Fe-O), and 224 formula unit (001) system (four layers
of O-Fe-O-Fe-O)
as starting configurations.  The vacuum configurations were then placed
in contact with a layer of water 
(about 1.5 nm thick) obtained from a simulation of bulk water
at 300 K.  Simulations of both the (012) and (001) interfacial
systems were run for 50 picoseconds 
at 300 K.  This time was deemed sufficiently long; by about 40 
picoseconds, the potential energy had ceased drifting downward and began
to fluctuate about its mean value.  

The surface structures for the (012) and (001) systems are 
shown in Figures \ref{012.solv} and \ref{001.solv}, respectively. 
For the (012) surface, as might have been expected, each bare iron
ion has indeed adsorbed a water molecule
from the solvent.  In several cases, ``chemisorption'' has occurred, and
some of the Fe$_3$O sites on the surface have protonated. 

For the (001) surface, however, the assumption of six-fold iron coordination
does not appear to be strictly valid; most iron atoms
are four-fold coordinated.  Here there is a balance
among the surface relaxation
forces (acting to pull the iron ion inward toward the surface), 
the forces of solvation (acting to pull the iron ion out towards the solution), and
the lateral surface hydrogen bonding forces.
Those surface iron ions which are 6-fold coordinated have slightly
``popped out'' of the relaxed top layer of iron ions.  Because the
relaxation energy for this surface is quite large, this evidently requires
a significant amount of work.  Thus the large surface relaxation energy
tends to keep the top iron layer in a four-fold coordinated state.

The issue of the coordination state of the top iron layer
is important because x-ray adsorption studies of the oxidation of
metallic iron have often revealed evidence for the presence of
four fold coordinated iron oxides in the passivation layer \cite{davenport95}.  
This has been used
to suggest that the passivation layer consists of maghemite ($\gamma$Fe$_2$O$_3$).  While this
seems likely due to the additional evidence from x-ray scattering, 
it is important to
note that the presence of four-fold coordination of iron does not 
appear to be a sufficient condition for the presence of maghemite.    

Of course, for a hematite crystal growing
in the presence of a supersaturated solution of iron ions, 
this artificially cleaved and hydroxylated surface will
adsorb iron ions and grow.  This study obviously says nothing
about the mechanisms of iron adsorption and crystal growth.  For
example, it is possible that our artificially-prepared surface has
an extremely high affinity for iron and would never be realized
for a surface in equilibrium with iron in solution.  In other words
it may only exist in nature over, say, microsecond 
time intervals (where no growth occurs)
after the solvation of a freshly fractured hematite crystal.
At the present time, however, it represents our best, least-biased guess 
for the structure of a hydroxylated hematite (001) surface.

These molecular dynamics 
studies are meant to reveal the coarse aspects of the
surface structure.  Interpretation beyond this is unwarranted because
the thermodynamic states of the systems during hydroxylation are not well-defined.
In particular, there is no reliable way of defining a solution pH, which is 
clearly a controlling variable in the extent of surface hydroxylation.
However, it is possible to make some fairly precise statements about
the predictions of the model concerning
the energetics of hydroxylation at 0 K with no solvent present at
the interface.  

\subsubsection{Structure and Energetics of Gas-phase Hydroxylation: Molecular Statics Calculations}
\bigskip
{\it (012) Hydroxylated Surface Structure}
\bigskip

For the (012) surface,
the first step toward ``gas phase'' surface hydroxylation is the adsorption of a 
water molecule to the five-fold coordinated iron ion (see Figure~\ref{012.hydrated}).
This produces a chain of water molecules sitting over the ``valleys'' formed
at the middle oxygens in the O-Fe-O-Fe-O sequence, spaced about 300 pm apart.  The
bond distance to the (formerly) five-fold coordinated iron ions is 212
pm. The  coordinates of atoms in  the unit cell of the slab for
hydrated and hydroxylated surface are given in Table~\ref{012hydrate}
and \ref{012hydrox}, respectively. 
As shown in Table~\ref{012relaxhydrox}, adsorption of water molecules has
only a minor effect on the rest of the surface.  Hydroxylation of the surface
takes place by dissociation of the adsorbed water and protonation of
the zig-zag ridge of oxygens (Figure~\ref{012.hydrox}).  The rows of
adsorbed hydroxide ions over the oxygen ``valleys'' 
move closer together (280 pm).  Hydroxylation has a significant
effect on the surface structure, moving the top layer of iron
ions much closer to the top of the oxygen ``ridges''.
The bond between the top-layer iron ion and the bottom-layer oxygen ion
in the O-Fe-O-Fe-O sequence is elongated from 210 pm (nonhydrated) to 212 pm
(hydrated) to 238 pm (hydroxylated).

Concerning the speciation of surface hydroxyls, per unit cell, there are
two $\equiv$Fe$_3$OH and two $\equiv$Fe$_1$OH sites, with a relatively
weak 203 pm hydrogen bond forming between the singly and triply
coordinated hydroxyls.  The O-O distance between $\equiv$Fe$_3$OH
and $\equiv$Fe$_1$OH is 290 pm.  The singly-coordinated sites are
nearly colinear along the X direction, spaced 280 pm apart.
Overall, the arrangement of $\equiv$Fe$_1$OH groups is
quite similar to the (110) surface of goethite in having a linear
array of singly-coordinated sites which accept hydrogen bonds from
triply-coordinated surface hydroxyls.  The main differences are that
the O-O distances are closer, the linear arrays are more closely
spaced in the Y direction (0.518 nm vs 1.139 nm) and the hydrogen
bonds donated by the triply coordinated hydroxyls are weaker.  The
hematite (012) surface should therefore prefer oxyanions with smaller
O-O distances and have a higher overall specific adsorptive capacity
than goethite (110).  Assume for the sake of argument that binding
is bidentate (eg. ($\equiv$FeOH)$_2$(PO$_4$) groups).
For
hematite (012) there are 2 bidentate sites per unit cell (0.288 nm$^2$) or
6.9 bidentate sites/nm$^2$,
whereas for goethite there is one bidentate site per unit cell (.358 nm$^2$),
or 2.8 bidentate sites/nm$^2$.
Also, because of the close spacing of the hydroxide
rows, hematite might be less effective in binding 
elongated adsorbates.

\bigskip
{\it (001) Hydroxylated Surface Structure}
\bigskip

For the (001) surface, the first step toward hydroxylation is the adsorption
of a water molecule directly over the three-fold coordinated iron ion (Figure~\ref{001.hydrated}).
For the minimized configuration, this breaks the symmetry and a rectangular unit
mesh results.  The adsorbed water molecule sits 207 pm above the (formerly) three-fold
coordinated iron ion. The  coordinates of atoms in  the unit cell of the slab for
hydrated and hydroxylated surface are given in Table~\ref{001hydrate}
and \ref{001hydrox}, respectively.  Again, as shown in
Table~\ref{001relaxhydrox}  
hydration of the (001) surface has almost no effect on
surface relaxation.  

Surface hydroxylation (Figure~\ref{001.hydrox}) takes place by dissociating the 
adsorbed water molecule and protonating 1/3 of the three-fold coordinated surface oxygens.
As for the (012) surface, hydroxylation has a significant effect on the
surface relaxation.  The top iron ion moves upward relative to the first oxygen layer
increasing the distance between the top Fe and O planes to approximately the bulk value
value.  The hydroxide layer is not indicated in Table~\ref{001relaxhydrox}
moves about 22 pm above the oxygen ions not receiving protons in the 
first structural oxygen layer.  There are no significant hydrogen bonding interactions
at the surface.

The surface hydroxyl speciation for (001) is two Fe$_3$OH and two FeOH sites
per unit cell (0.465 nm$^2$), resulting in a density of 4.3 sites/nm$^2$
of each type (seperately).  There is no possibility for bidentate binding
at the (001) surface, at least for the model arrived at here.

\bigskip
{\it Hydroxylation Energetics}
\bigskip

The
energy of adsorption of water to the (012) surface in the configuration
shown in Figure~\ref{012.hydrated} is -1.6 J/m$^2$ (the adsorbed water
molecule is assumed to be removed 
from bulk ice at 0 K).  If, in addition, the water is allowed to dissociatively
adsorb (as shown in Figure~\ref{012.hydrox}), the adsorption energy is -2.7 J/m$^2$.  Therefore, 
one obtains more energy for
adsorption of the water layer than the 2 J/m$^2$ required to fracture the crystal.  
This is clearly not correct, since hematite crystals are stable in the presence
of water. 

The same considerations apply to the ``gas phase'' hydroxylation of the (001) surface.
Here the energy of adsorption of undissociated water in the configuration shown in
Figure~\ref{001.hydrated} is -1.2 J/m$^2$.  The energy of dissociative adsorption
in the configuration shown in Figure~\ref{001.hydrox} is -1.8 J/m$^2$.  Again
the energy of
dissociative adsorption exceeds the surface fracture energy (1.6 J/m$^2$).

The surface enthalpy of hematite has been estimated 
in reference \cite{langmuir71} by measuring the heat of dissolution of
hematite crystals as a function of surface area \cite{ferrier66}.  A value
of +0.770 J/m$^2$ was estimated for the surface enthalpy at 70$^\circ$C.      
Taking this result at face value, and assuming the experimental value
is representative of the (012) surface, there is a discrepency of 1.4
J/m$^2$ between the experimental value and the theoretical estimate.
The failure to obtain the correct sign could be due to either
overestimating the binding energy of water, hydroxide, and protons to
the surface, or underestimating the energy required to fracture the surface.
As discussed above, nearly all theoretical 
estimates of the fracture energy are in
fair agreement
predicting that the surface energy is 1.6-2.0 J/m$^2$
yet experimental estimates range from 6-40 J/m$^2$.
The discrepency of 1.4 J/m$^2$ is small in relation to this 
range of experimental values for the fracture energy. 
The same arguments apply if the predominant surface in the surface
enthalpy experiments is (001).
Meaningful discussion of the hydroxylation issue must wait until
the overall discrepency between 
the theoretical and experimental fracture surface 
energies is better understood.  We will, however, point out that
our model predicts a greater amount of dissociation on (012) than (001),
both in terms of ``gas phase'' calculations and qualitatively in
the MD simulations of the hematite-water interface.

\section{Summary and Conclusions}

We have presented a simple yet rigorous derivation of the 2-D Ewald
summation method of Rhee et al.~\cite{rhee-halley-hautman-rahman89}
and its generalization to
polarizable systems. This provides a viable means of investigating
the surfaces of oxide minerals using thousands of
atoms.  As one can see from the presented results
of benchmarks against exact methods,
the method of Rhee and co-workers is adequate
for systems with aspect ratios ($(z_{max}-z_{min})/min(L_x,L_y)$) less
than approximately 0.5.  We increased the order of the Taylor
expansion for V$_m$ from 4 (as in \cite{rhee-halley-hautman-rahman89}
and \cite{hautman-klein92}) and found this to yield significantly
better accuracy.  These expressions are given in the Appendix.

We have applied the method to investigate the structures of
hydroxylated and non-hydroxylated (012) and (001) surfaces
of hematite ($\alpha$-Fe$_2$O$_3$) in slab geometry, and solidly
document the convergence of our results
with respect to slab thickness.  Our nonhydroxylated
structures and energies are in good agreement with previous calculations
on isostructural $\alpha$-Al$_2$O$_3$.  We used the flexibility
and economy of our parameterized analytical potential approach
to run simulations of solvated interfaces.  These simulations
revealed nonuniform hydroxylation as suggested in \cite{brown90},
but revealed end-member structures which were used in more
precise ``gas-phase'' calculations.  These calculations revealed
that the effect of hydration on surface structure is small, but
that hydroxylation has a significant effect.  Hydroxylation
appears to be energetically favored both in the gas-phase
calculations and the solution phase simulations, with the
(012) surface exhibiting a greater degree of dissociation
of adsorbed water.  The (001) surface has a complex
solvated structure.  Our simulations reveal the presence
of both four-fold coordinated and six-fold coordinated iron
ions at the solvated surface, with four-fold coordination
predominant.  Conversion of four-fold to six-fold coordination
requires significant movement of top-layer iron atoms
out into the aqueous solvent.  This movement should be detected
in x-ray reflectivity experiments designed to measure the
location of the top layer of iron ions and their bound
hydroxyl/aquo ligands in the direction normal to the surface
plane \cite{als-nielsen91,chiarello-sturchio95}.  It is
hoped that the present study will encourage experimental
studies of hydroxylated hematite surfaces.

Quantum mechanical simulations are potentially superior to
those with parameterized analytical potentials because
electronic structure is accounted for explicitly.  However,
these methods must always compromise some accuracy
for the sake of feasibility.  This means that even
demonstrating convergence with respect to basis set size,
density of $k$-points mesh for Brillouin zone integration, system
size requires tremendous effort.  As shown in Tables~\ref{quantum001_e}
and~\ref{quantum001_z}, results of electronic structure
based methods are not always in agreement.  We therefore
wish to emphasize the usefulness of parameterized analytical
potential methods in extending the scope of theoretical
calculations on mineral surfaces.

\section*{Acknowledgement}
The authors thank Dr. Greg Schenter of the Pacific Northwest National
Laboratory for providing his code that implements the Parry method for
the benchmark calculations.  JRR is grateful to the Minnesota
Supercomputer Institute for hospitality over the course of completing
this work and to Dr. Neil C. Sturchio of Argonne National Laboratory
for the discussion of experimental techniques for surface structure
analysis.  This work was supported by the Office of Basic Energy
Sciences, Engineering and Geosciences Division, contract 18328, and by
Laboratory Directed Research and Development funds from Pacific
Northwest National Laboratory. Pacific Northwest National Laboratory is
operated for the U.S. Department of Energy by Battelle Memorial
Institute under Contract DE-AC06-76RL0 1830.


\appendix
\section*{MULTIPOLE EXPANSIONS}
The first, second, and third derivatives of V$_m$ in
equation~\ref{vm} are required for computation of
energies and forces for charge-dipole and dipole-dipole
interactions in equations~\ref{phicd}~and~\ref{phidd}.
\begin{eqnarray}
{{\partial v_m^{\prime}}\over{\partial x}} & = & {{3}\over{2}}
z^2 x \sum_{\nu=\nu_{\rm max}}^\infty{{1}\over{\nu^5}}
-{{15}\over{2}} z^2 x \sum_{\nu=\nu_{\rm
max}}^\infty{{\nu_x^2}\over{\nu^7}}
-{{15}\over{4}} s^2 z^2 x \sum_{\nu=\nu_{\rm
max}}^\infty{{1}\over{\nu^7}}
-{{15}\over{8}} z^4 x \sum_{\nu=\nu_{\rm
max}}^\infty{{1}\over{\nu^7}}\nonumber\\
&&-{{105}\over{4}} z^2 x (s^2  \sum_{\nu=\nu_{\rm
max}}^\infty{{\nu_x^2}\over{\nu^9}} 
+\sum_{\nu=\nu_{\rm max}}^\infty{{{(\vec
s\cdot\vec\nu)}^2}\over{\nu^9}})
+{{105}\over{8}} z^4 x \sum_{\nu=\nu_{\rm
max}}^\infty{{\nu_x^2}\over{\nu^9}}\\
{{\partial v_m^{\prime}}\over{\partial z}} & = & -z\sum_{\nu=\nu_{\rm
max}}^\infty{{1}\over{\nu^3}} 
+{{3}\over{2}}s^2 z \sum_{\nu=\nu_{\rm max}}^\infty{{1}\over{\nu^5}}
+{{3}\over{2}} z^3 \sum_{\nu=\nu_{\rm
max}}^\infty{{1}\over{\nu^5}}\nonumber\\
&&-{{15}\over{2}} z\sum_{\nu=\nu_{\rm max}}^\infty {{{(\vec s \cdot
\vec\nu)}^2}\over{\nu^7}}
-{{15}\over{8}} s^4 z \sum_{\nu=\nu_{\rm max}}^\infty{{1}\over{\nu^7}}
-{{15}\over{4}} s^2 z^3 \sum_{\nu=\nu_{\rm max}}^\infty{{1}\over{\nu^7}}
-{{15}\over{8}} z^5 \sum_{\nu=\nu_{\rm
max}}^\infty{{1}\over{\nu^7}}\nonumber\\
&&+{{105}\over{4}} s^2 z \sum_{\nu=\nu_{\rm max}}^\infty{{{(\vec s \cdot
\vec\nu)}^2}\over{\nu^9}}
+{{105}\over{4}} z^3 \sum_{\nu=\nu_{\rm max}}^\infty{{{(\vec s \cdot
\vec\nu)}^2}\over{\nu^9}}\\
{{\partial^2 v_m^\prime}\over{\partial x^2}}&=&{{3}\over{2}} z^2
\sum_{\nu=\nu_{\rm max}}^\infty{{1}\over{\nu^5}}
-{{15}\over{2}} z^2 \sum_{\nu=\nu_{\rm
max}}^\infty{{\nu_x^2}\over{\nu^7}}
-{{15}\over{4}} z^2(s^2+2 x^2) \sum_{\nu=\nu_{\rm
max}}^\infty{{1}\over{\nu^7}}
-{{15}\over{8}} z^4 \sum_{\nu=\nu_{\rm
max}}^\infty{{1}\over{\nu^7}}\nonumber\\
&&+{{105}\over{4}} z^2(s^2+4x^2)\sum_{\nu=\nu_{\rm
max}}^\infty{{\nu_x^2}\over{\nu^9}} 
+{{105}\over{8}} z^4 \sum_{\nu=\nu_{\rm
max}}^\infty{{\nu_x^2}\over{\nu^9}}
+{{105}\over{4}} z^2 \sum_{\nu=\nu_{\rm max}}^\infty{{{(\vec s \cdot
\vec\nu)}^2}\over{\nu^9}}\\
{{\partial^2 v_m^{\prime}} \over{\partial x\partial y}}&=&
{{-15}\over{2}}z^2 y x \sum_{\nu=\nu_{\rm max}}^\infty{{1}\over{\nu^7}}+
{{105}\over{2}}z^2 y x(\sum_{\nu=\nu_{\rm
max}}^\infty{{\nu_x^2}\over{\nu^9}}+ 
\sum_{\nu=\nu_{\rm max}}^\infty{{\nu_y^2}\over{\nu^9}})\\
{{\partial^2 v_m^{\prime}}\over{\partial x\partial
z}}&=&3xz\sum_{\nu=\nu_{\rm max}}^\infty{{1}\over{\nu^5}}
-{{15}\over{2}}s^2xz \sum_{\nu=\nu_{\rm max}}^\infty{{1}\over{\nu^7}}
-{{15}\over{2}}z^3x \sum_{\nu=\nu_{\rm max}}^\infty{{1}\over{\nu^7}}
-15zx \sum_{\nu=\nu_{\rm
max}}^\infty{{\nu_x^2}\over{\nu^7}}\nonumber\\
&&+{{105}\over{2}}zx(s^2\sum_{\nu=\nu_{\rm
max}}^\infty{{\nu_x^2}\over{\nu^9}}
+\sum_{\nu=\nu_{\rm max}}^\infty{{{(\vec s \cdot
\vec\nu)}^2}\over{\nu^9}})
+{{105}\over{2}}z^3x\sum_{\nu=\nu_{\rm
max}}^\infty{{\nu_x^2}\over{\nu^9}}\\
{{\partial^2 v_m^{\prime}}\over{\partial z^2}}&=&-\sum_{\nu=\nu_{\rm
max}}^\infty{{1}\over{\nu^3}} 
+{{3}\over{2}}(s^2+3z^2) \sum_{\nu=\nu_{\rm
max}}^\infty{{1}\over{\nu^5}}
-{{15}\over{2}} \sum_{\nu=\nu_{\rm max}}^\infty {{{(\vec s \cdot
\vec\nu)}^2}\over{\nu^7}}\nonumber\\
&&-{{15}\over{8}} (s^4+6s^2 z^2+5z^5) \sum_{\nu=\nu_{\rm
max}}^\infty{{1}\over{\nu^7}}
+{{105}\over{4}} s^2 z \sum_{\nu=\nu_{\rm max}}^\infty{{{(\vec s \cdot
\vec\nu)}^2}\over{\nu^9}}\nonumber\\
&&+{{105}\over{4}} (s^2+3z^2) \sum_{\nu=\nu_{\rm max}}^\infty{{{(\vec s
\cdot \vec\nu)}^2}\over{\nu^9}}\\
{{\partial^3 v_m^{\prime}}\over{\partial x^3}}&=&6x\left({{-
15}\over{4}}z^2\sum_{\nu=\nu_{\rm max}}^\infty{{1}\over{\nu^7}}
+{{105}\over{2}}z^2\sum_{\nu=\nu_{\rm
max}}^\infty{{\nu_x^2}\over{\nu^9}}\right)\\
{{\partial^3 v_m^\prime}\over{\partial x^2\partial y}}&=&{{-
15}\over{2}}z^2y\sum_{\nu=\nu_{\rm max}}^\infty{{1}\over{\nu^7}}
+{{105}\over{2}}z^2 y\left(\sum_{\nu=\nu_{\rm
max}}^\infty{{\nu_x^2}\over{\nu^9}}+
\sum_{\nu=\nu_{\rm max}}^\infty{{\nu_y^2}\over{\nu^9}} \right)\\
{{\partial^3 v_m^{\prime}}\over{\partial x\partial y\partial z}}&=&-
15xyz\sum_{\nu=\nu_{\rm max}}^\infty{{1}\over{\nu^7}}
+105xyz\left(\sum_{\nu=\nu_{\rm max}}^\infty{{\nu_x^2}\over{\nu^9}}+
\sum_{\nu=\nu_{\rm max}}^\infty{{\nu_y^2}\over{\nu^9}} \right)\\
{{\partial^3 v_m^{\prime}}\over{\partial x^2\partial
z}}&=&3z\sum_{\nu=\nu_{\rm max}}^\infty{{1}\over{\nu^5}}
-15z\sum_{\nu=\nu_{\rm max}}^\infty{{\nu_x^2}\over{\nu^7}}
-{{15}\over{2}} (s^2+2 x^2) \sum_{\nu=\nu_{\rm
max}}^\infty{{1}\over{\nu^7}}
-{{15}\over{2}} z^3 \sum_{\nu=\nu_{\rm
max}}^\infty{{1}\over{\nu^7}}\nonumber\\
&&+{{105}\over{2}} z(s^2+4x^2)\sum_{\nu=\nu_{\rm
max}}^\infty{{\nu_x^2}\over{\nu^9}} 
+{{105}\over{2}} z^3 \sum_{\nu=\nu_{\rm
max}}^\infty{{\nu_x^2}\over{\nu^9}}
+{{105}\over{2}} z \sum_{\nu=\nu_{\rm max}}^\infty{{{(\vec s \cdot
\vec\nu)}^2}\over{\nu^9}}\\
{{\partial^3 v_m^{\prime}}\over{\partial x\partial
z^2}}&=&3x\sum_{\nu=\nu_{\rm max}}^\infty{{1}\over{\nu^5}}
-{{15}\over{2}}s^2x \sum_{\nu=\nu_{\rm max}}^\infty{{1}\over{\nu^7}}
-{{45}\over{2}}z^2x \sum_{\nu=\nu_{\rm max}}^\infty{{1}\over{\nu^7}}
-15x \sum_{\nu=\nu_{\rm
max}}^\infty{{\nu_x^2}\over{\nu^7}}\nonumber\\
&&+{{105}\over{2}}x(s^2\sum_{\nu=\nu_{\rm
max}}^\infty{{\nu_x^2}\over{\nu^9}}
+\sum_{\nu=\nu_{\rm max}}^\infty{{{(\vec s \cdot
\vec\nu)}^2}\over{\nu^9}})
+{{315}\over{2}}z^2x\sum_{\nu=\nu_{\rm
max}}^\infty{{\nu_x^2}\over{\nu^9}}\\
{{\partial^3 v_m^{\prime}}\over{\partial z^3}}&=&+9z \sum_{\nu=\nu_{\rm
max}}^\infty{{1}\over{\nu^5}}
-{{45}\over{2}} s^2z \sum_{\nu=\nu_{\rm max}}^\infty{{1}\over{\nu^7}}
-{{75}\over{2}} z^3 \sum_{\nu=\nu_{\rm max}}^\infty{{1}\over{\nu^7}}
+{{315}\over{2}} z \sum_{\nu=\nu_{\rm max}}^\infty{{{(\vec s \cdot
\vec\nu)}^2}\over{\nu^9}}
\end{eqnarray}


\begin{figure}
\caption{Nonhydrated (012) surface.  (a) slab cross section
of the system consisting of 3 layers of neutral O-Fe-O-Fe-O sequences.
Large red atoms are oxygen,
small orange atoms are iron. (b-d) closeup views of top O-Fe-O-Fe-O layer
plus the top O layer from the middle neutral stacking sequence.
Atoms are color coded according to depth.  The top O layer is red,
the middle O layer is magenta, and the bottom O layer is pink.  The
top O layer of the middle stacking sequence is grey.  The top five-fold
coordinated iron ions are brown, and the lower six-fold coordinated iron
ions are orange.  In the map view (d) the zig-zag oxygen ridges are
emphasized with black vectors.}
\label{012.nohydrox}
\end{figure}


\begin{figure}
\caption{
Nonhydrated (001) surface.  (a) slab cross section of the system
consisting of four layers of neutral Fe-O-Fe sequences.  Large red
atoms are oxygen, small orange atoms are iron. (b-d) closeup views of
top two Fe-O-Fe layers but minus the bottom Fe layer from the 2nd
neutral Fe-O-Fe stacking sequence.  Atoms are color coded according to
depth.  The top O layer is red, and the bottom O layer is deep pink.
The top three-fold coordinated ion ions are brown, the second iron
layer is orange and the third iron layer is grey.
}
\label{001.nohydrox}
\end{figure}


\begin{figure}
\caption{
A snapshot (after 50 picoseconds) of the solvated (012)
surface.  The initial conditions of the simulation consisted were
obtained by placing a 1.5 nm layer of water from a bulk water
simulation over the surface shown in Figure~\protect\ref{012.nohydrox}(a).
Atom types are the same as in Figure~\protect\ref{012.nohydrox} except
that both surface water molecules and solvent hydroxide ions are
indicated by large yellow oxygen atoms.  A water molecule is
considered a surface water molecule if it is bound to at least 1 iron
atom, and a hydroxide ion is considered solvated if it is not bound to
any iron atoms.
}
\label{012.solv}
\end{figure}


\begin{figure}
\caption{A snapshot (after 50 picoseconds) of the solvated (001)
surface.  The initial conditions of the simulation consisted were
obtained by placing a 1.5 nm layer of water from a bulk water
simulation over the surface shown in
Figure~\protect\ref{001.nohydrox}(a).  Atom types are the same as in
Figure~\protect\ref{001.nohydrox} except that both surface water
molecules and solvent hydroxide ions are indicated by large yellow
oxygen atoms.  A water molecule is considered a surface water molecule
if it is bound to at least 1 iron atom, and a hydroxide ion is
considered solvated if it is not bound to any iron atoms.}
\label{001.solv}
\end{figure}

\begin{figure}
\caption{Three orthogonal views of the hydrated (012) surface.
Atom types are the same as in Figure~\protect\ref{012.nohydrox} except that
adsorbed water molecules are indicated by the yellow oxygen
atoms and protons are the small white
atoms.  Idealized periodic surface structures are those suggested
from the MD simulations shown in Figure~\protect\ref{012.solv}}
\label{012.hydrated}
\end{figure}


\begin{figure}
\caption{
Three orthogonal views of the hydroxylated (012) surface.
Atom types are the same as in Figure~\protect\ref{012.nohydrox} except that
adsorbed hydroxide ions are indicated by the yellow oxygen
atoms and protons are the small white
atoms.  Idealized periodic surface structures are those suggested
from the MD simulations shown in Figure~\protect\ref{012.solv}}
\label{012.hydrox}
\end{figure}

\begin{figure}
\caption{
Three orthogonal views of the hydrated (001) surface.
Atom types are the same as in Figure~\protect\ref{001.nohydrox} except that
adsorbed water molecules are indicated by the yellow oxygen
atoms and protons are the small white
atoms.  Idealized periodic surface structures are those suggested
from the MD simulations shown in Figure~\protect\ref{001.solv}}
\label{001.hydrated}
\end{figure}


\begin{figure}
\caption{
Three orthogonal views of the hydroxylated (001) surface.
Atom types are the same as in Figure~\protect\ref{012.nohydrox} except that
adsorbed hydroxide ions are indicated by the yellow oxygen
atoms and protons are the small white
atoms.  Idealized periodic surface structures are those suggested
from the MD simulations shown in Figure~\protect\ref{001.solv}}
\label{001.hydrox}
\end{figure}


\begin{table}
\caption{The energy (in units of $e^2/L$) and
 forces (in units of $e^2/L^2$),between a charge $+e$ at (0, 0, 0)
and $-e$ at (0.1, 0.1, 0.1) L  in an infinitely periodic 2-D array of
square  unit cells}
\label{bench1} 
\begin{tabular}{ccccc}
Method & U &    $F_x$  &  $F_y$ &  $F_z$ \\
Leckner $(l_{max}=10)$ &  -5.77211 & - & - & - \\
Parry ($\kappa=5$) & -5.77212 &18.8136 & 18.8136 & 20.1628 \\
This study & -5.77184 & 18.8136 & 18.8136 & 20.1627 \\
\end{tabular}
\end{table}

\begin{table}
\caption{The energy (in units of $e^2/L$) and
 forces (in units of $e^2/L^2$),between a charge $+e$ at (0, 0, 0)
and $-e$ at (-0.25, -0.15, -0.2)L  in an infinitely periodic 2-D array of 
square  unit cells}
\label{bench2} 
\begin{tabular}{ccccc}
Method                & U   &   $F_x$  &  $F_y$ &  $F_z$ \\ 
Leckner $(l_{max}=10)$ & -2.82156 & -4.60495 & -2.91052 & -6.44748 \\
Parry ($\kappa=5$)     & -2.82157 & -4.60495 & -2.91052 & -6.447478 \\
This study ($\gamma=0.1 L_{short})$ &  -2.8214336 & -4.60323 & -2.909703 
& -6.446037 \\ 
\end{tabular}
\end{table}
\begin{table}
\caption{The energy (in units of $e^2/L$) and
 forces (in units of $e^2/L^2$),between a charge $+e$ at (0, 0, 0)
and $-e$ at (-0.25, -0.15, -0.5)L in an infinitely periodic 2-D array of
square  unit cells}
\label{bench3} 
\begin{tabular}{ccccc}
Method & U &    $F_x$  &  $F_y$ &  $F_z$ \\
Parry's & -0.8061173 & -0.6608550 & -0.4514618 & -6.723419\\
This study & -0.8067679 & -0.650034 & -0.4461202 & -6.5550016 \\
Error, \%   & 0.08 & 1.64 & 1.18 & 2.50 \\
\end{tabular}
\end{table}
\begin{table}
\caption{The energy (in units of $e^2/\AA$) and
 forces (in units of $e^2/\AA^2$),between a charge $+e$ at (0, 0, 0)
and $-e$ at (-0.25, -0.15, -0.1)L and also dipole 
$\vec \mu=(-0.33, -0.16, 0.77 ) $ at the position of $-e$ 
in an infinitely 
periodic 2-D array of
square unit cells with the side length $L=10$ \AA }
\label{bench4} 
\begin{tabular}{ccccc}
Method & U &    $F_x\times 10^2$ &  $F_y \times 10^2$ &  $F_z \times 10^2$ \\
Parry's & -0.33385337 & -7.280129828 & -4.56935990 & -8.077974417 \\
This study & -0.336207073 & -7.2792028 & -4.56935929 & -8.077045 \\
\end{tabular}
\end{table}
\begin{table}
\caption{The energy (in units of $e^2/\AA$) and force 
 (in units of $e^2/\AA^2$) in a system consisting of a charge $+2e$ at
(0,0,0), a charge $-e$ and dipole $\vec \mu = (-0.33, -0.16, 0.77)$ at
$(-0.25, -0.15, -0.1)L$ and a charge -e and dipole $\vec \mu=(0.3, 0.6,
0.7)$ at $(0.5,0.33,-0.05)L$ in an infinitely periodic 2-D array of
square unit cells with the side length $L=10$ \AA }

\label{bench_dd} 
\begin{tabular}{ccccc}
Method            & U   &   $F_x$  &  $F_y$ &  $F_z$ 
\protect\tablenote{Only the force acting on charge $+2e$ is given} \\
Parry's & -0.8925432834 & -0.13797318 & -0.062559269  &  -0.2051223 \\
This study & -0.89619000 & -0.13788360  & -0.0625851709 & -0.20480698 \\
\end{tabular}
\end{table}

\begin{table}
\caption{ Slab unit cell coordinates for the ideal vacuum-terminated (012) surface.  
X, Y, and Z coordinates in \AA .   
Orthogonal unit mesh translation vectors are 5.564 \AA\ in X and
5.176 \AA\ in Y. }  
\label{012vacuum}
\begin{tabular}{cccc}
atom&x&y&z\\
O1    &   0.374  &   0.254  &   1.169  \\
O2    &   2.689  &   2.746  &   4.819  \\
O3    &   3.605  &   2.856  &  -2.573  \\
O4    &   2.782  &   0.000  &  -3.870  \\
O5    &   2.066  &   0.000  &   3.868  \\
O6    &   5.206  &   2.608  &  -0.001  \\
O7    &   2.160  &   2.746  &  -4.820  \\
O8    &   1.244  &   2.857  &   2.572  \\
O9    &   4.474  &   0.254  &  -1.170  \\
O10   &   1.692  &   4.432  &  -1.171  \\
O11   &   4.942  &   1.940  &  -4.820  \\
O12   &   4.026  &   1.830  &   2.572  \\
O13   &   4.848  &   4.687  &   3.868  \\
O14   &   2.424  &   2.078  &  -0.001  \\
O15   &   0.000  &   4.686  &  -3.870  \\
O16   &   0.823  &   1.830  &  -2.573  \\
O17   &   5.471  &   1.941  &   4.819  \\
O18   &   3.156  &   4.432  &   1.169  \\
Fe1   &   0.901  &   1.056  &  -0.786  \\
Fe2   &   3.962  &   3.526  &  -4.531  \\
Fe3   &   3.332  &   3.656  &   3.028  \\
Fe4   &   1.516  &   3.655  &  -3.029  \\
Fe5   &   0.887  &   3.526  &   4.530  \\
Fe6   &   3.947  &   1.056  &   0.785  \\
Fe7   &   3.669  &   1.161  &   4.530  \\
Fe8   &   1.165  &   3.630  &   0.785  \\
Fe9   &   4.298  &   1.030  &  -3.029  \\
Fe10  &   1.180  &   1.160  &  -4.531  \\
Fe11  &   0.550  &   1.031  &   3.028  \\
Fe12  &   3.683  &   3.630  &  -0.786  \\
\end{tabular}
\end{table}

\begin{table}
\caption{Changes in interplanar spacing due to
relaxation of hematite (012) surface as a function of size of the
simulation cell}
\label{012relaxstr}
\begin{tabular}{ccccc}
&& \multicolumn{3}{c}{Change from bulk value} \\ 
&& \multicolumn{3}{c}{Number of Fe$_2$O$_3$ molecules}\\
&Bulk value (pm) & 144&216&288\\
O&&&\\
&35 & -24\% &-18\% &-18\%\\
Fe&&&\\
&80&-24\%&-17\%&-18\%\\
O&&&\\
&80&8\%&5\%&5\%\\
Fe&&&\\
&35&49\%&32\%&34\%\\
O&&&\\
&149&-15\%&-6\%&-7\%\\
O&&&\\
\end{tabular}
\end{table}

\begin{table}
\caption{Surface energy $\gamma$ (J/m$^2$) for non-hydroxylated (012)
face of hematite calculated using the 3-D and 2-D periodic boundary
conditions}
\label{gamma012}
\begin{tabular}{cccc}
Number of Fe$_2$O$_3$ molecules & $\gamma_{3-D}$ relaxed & $\gamma_{2-D}$ relaxed & $\gamma_{3-D}$ unrelaxed \\
72  & 2.335  & 2.325  & 2.903 \\
144 & 1.882  & 1.874  & 2.757 \\
216 & 1.994  & 2.022  & 2.775 \\
288 & 1.999  & 2.094  & 2.787 \\
\end{tabular}
\end{table}

\begin{table}
\caption{Slab unit cell coordinates for the ideal vacuum-terminated
(001) surface.    
X, Y, and Z coordinates in \AA .   
Orthogonal unit mesh translation vectors are 8.976 \AA\ in X and
5.176 \AA\  in Y.}   
\label{001vacuum}
\begin{tabular}{cccc}
atom&x&y&z\\
O1    &  0.000 &  4.967 &  -1.188 \\
O2    &  1.410 &  2.409 &   3.428 \\
O3    &  4.488 &  2.380 &  -1.188 \\
O4    &  5.898 &  4.997 &   3.428 \\
O5    &  7.360 &  2.497 &   3.428 \\
O6    &  5.847 &  0.000 &  -1.188 \\
O7    &  1.359 &  2.587 &  -1.188 \\
O8    &  2.872 &  5.084 &   3.428 \\
O9    &  8.798 &  5.011 &   3.428 \\
O10   &  7.231 &  2.365 &  -1.188 \\
O11   &  2.743 &  4.952 &  -1.188 \\
O12   &  4.310 &  2.423 &   3.428 \\
O13   &  4.317 &  0.752 &  -3.428 \\
O14   &  5.727 &  3.369 &   1.188 \\
O15   &  8.805 &  3.340 &  -3.428 \\
O16   &  1.239 &  0.782 &   1.188 \\
O17   &  8.856 &  3.162 &   1.188 \\
O18   &  7.343 &  0.665 &  -3.428 \\
O19   &  2.855 &  3.252 &  -3.427 \\
O20   &  4.368 &  0.574 &   1.188 \\
O21   &  5.905 &  3.326 &  -3.428 \\
O22   &  7.472 &  0.796 &   1.188 \\
O23   &  1.418 &  0.738 &  -3.428 \\
O24   &  2.984 &  3.384 &   1.188 \\
Fe1   &  4.359 &  4.167 &  -3.863 \\
Fe2   &  5.856 &  1.580 &   0.290 \\
Fe3   &  8.847 &  1.580 &  -3.863 \\
Fe4   &  1.368 &  4.168 &   0.290 \\
Fe5   &  5.855 &  1.580 &  -2.603 \\
Fe6   &  7.352 &  4.168 &   2.209 \\
Fe7   &  1.367 &  4.168 &  -2.602 \\
Fe8   &  2.864 &  1.581 &   2.209 \\
Fe9   &  8.847 &  1.581 &  -0.289 \\
Fe10  &  5.856 &  1.582 &   3.864 \\
Fe11  &  1.368 &  4.169 &   3.864 \\
Fe12  &  4.359 &  4.169 &  -0.289 \\
Fe13  &  8.848 &  1.581 &   2.603 \\
Fe14  &  7.351 &  4.168 &  -2.209 \\
Fe15  &  2.864 &  1.581 &  -2.209 \\
Fe16  &  4.360 &  4.168 &   2.603 \\ 
\end{tabular}
\end{table}
\begin{table}
\caption{Changes in  interplanar spacing due to
relaxation of hematite (001) surface as a function of size of the
simulation cell}
\label{001relaxstr}
\begin{tabular}{cccccc}
&& \multicolumn{4}{c}{Change from bulk value} \\ 
&& \multicolumn{4}{c}{Number of Fe$_2$O$_3$ molecules}\\
&Bulk value (pm) &168&224&280&336\\
Fe&&&&&\\
&86 &-53\%& -50\% &-49\% & -49\%\\
O&&&&&\\
&86&-3\%&-3\%&-2\%&-3\%\\
Fe&&&&\\
&63&-43\%&-38\%&-43\%&-41\%\\
Fe&&&&\\
&86&24\%&19\%&21\%&21\%\\
O&&&&\\
\end{tabular}
\end{table}

\begin{table}
\caption{Surface Energies of Non-hydroxylated Fe/Al Sesquioxide (J/m$^2$):
Comparison with previous electronic structure calculations}
\label{quantum001_e} 
\begin{tabular}{ccccc}
&$\alpha$-Fe$_2$O$_3$& $\alpha$-Al$_2$O$_3$ &\\
&Ours&LDA \cite{manassidis-devita-gillan93}
&HF\cite{mackrodt92}&HF\cite{causa-dovesi-pisani-roetti89}  \\ 
(001)& & & &\\
Unrelaxed&4.24&3.77&3.3&5.0\\
Relaxed&1.64&1.76&2.0&-\\
(012)& & & &\\
Unrelaxed&2.79&2.95&-&-\\
Relaxed&2.00&-&-&-\\
\end{tabular}
\end{table}

\begin{table}
\caption{Surface energy $\gamma$ (J/m$^2$) for 
non-hydroxylated (001) face of hematite
calculated using the 3-D and 2-D periodic boundary conditions}
\label{gamma001}
\begin{tabular}{cccc}
Number of Fe$_2$O$_3$ molecules & $\gamma_{3-D}$ relaxed & $\gamma_{2-D}$ relaxed & $\gamma_{3-D}$ unrelaxed \\
112 & 1.63 & 1.62 & 4.32 \\
168 & 1.54 & 1.46 & 4.22 \\
224 & 1.64 & 1.63 & 4.22 \\
280 & 1.64 & 1.64 & 4.23 \\
336 & 1.65 & 1.60 & 4.24 \\
\end{tabular}
\end{table}

\begin{table}
\caption{Nonhydroxylated (001) surface. Percent relaxation in $z$ direction}
\label{quantum001_z} 
\begin{tabular}{cccccc}
  &Ours&LDA \protect\cite{manassidis-devita-gillan93}&TBTE 
\protect\cite{godin-lafemina94} 
&HF\cite{mackrodt92}&HF\cite{causa-dovesi-pisani-roetti89}  \\ 
Fe/Al&    &     &  &  \\
  & -49&-86  &-87  &-50&-50  \\
O &    &     &  &  \\
  & -3 &+3   &-33  & -5&- \\
Fe/Al&    &     &  &  & \\ 
  & -41&-54  &+20  & -8 &-\\
Fe/Al&    &     &  &  &\\ 
  &  21&25   &0  &- &- \\
O &    &     &  &  & \\
\end{tabular}
\end{table}

\begin{table}
\caption{Slab unit cell coordinates for the hydrated (012) surface.  
X, Y, and Z coordinates in \AA .   
Orthogonal unit mesh translation vectors are 5.564 \AA\  in X and
5.176 \AA\ in Y.}   
\label{012hydrate}
\begin{tabular}{cccc}
atom&x&y&z\\
H1   &  1.856 &  0.028 & -6.877  \\
H2   &  1.122 &  1.264 & -7.312  \\
H3   &  3.728 &  1.264 &  7.314  \\
H4   &  2.987 &  0.033 &  6.879  \\
H5   &  0.946 &  3.942 &  7.312  \\
H6   &  0.221 &  0.000 &  6.883  \\
H7   &  4.628 &  0.003 & -6.882  \\
H8   &  3.906 &  3.938 & -7.313  \\
O1   &  1.135 &  0.643 & -6.531  \\
O2   &  0.377 &  0.519 &  1.166  \\
O3   &  2.721 &  3.024 &  4.809  \\
O4   &  3.596 &  3.123 & -2.575  \\
O5   &  2.782 &  0.264 & -3.846  \\
O6   &  2.068 &  0.264 &  3.846  \\
O7   &  5.207 &  2.873 &  0.001  \\
O8   &  3.712 &  0.644 &  6.533  \\
O9   &  2.129 &  3.024 & -4.809  \\
O10  &  1.253 &  3.123 &  2.576  \\
O11  &  4.472 &  0.520 & -1.165  \\
O12  &  0.931 &  4.563 &  6.532  \\
O13  &  1.691 &  4.690 & -1.165  \\
O14  &  4.910 &  2.185 & -4.809  \\
O15  &  4.035 &  2.086 &  2.576  \\
O16  &  4.849 &  4.945 &  3.847  \\
O17  &  2.425 &  2.337 &  0.000  \\
O18  &  0.000 &  4.945 & -3.845  \\
O19  &  0.815 &  2.086 & -2.575  \\
O20  &  5.503 &  2.186 &  4.810  \\
O21  &  3.923 &  4.560 & -6.533  \\
O22  &  3.159 &  4.690 &  1.166  \\
Fe1  &  0.902 &  1.315 & -0.786  \\
Fe2  &  3.957 &  3.813 & -4.546  \\
Fe3  &  3.332 &  3.908 &  3.022  \\
Fe4  &  1.518 &  3.908 & -3.021  \\
Fe5  &  0.893 &  3.813 &  4.547  \\
Fe6  &  3.948 &  1.315 &  0.787  \\
Fe7  &  3.675 &  1.396 &  4.548  \\
Fe8  &  1.166 &  3.894 &  0.787  \\
Fe9  &  4.300 &  1.302 & -3.021  \\
Fe10 &  1.175 &  1.396 & -4.546  \\
Fe11 &  0.550 &  1.301 &  3.021  \\
Fe12 &  3.684 &  3.894 & -0.786  \\  
\end{tabular}
\end{table}
\begin{table}
\caption{Slab unit cell coordinates for the hydroxylated (012) surface.  
X, Y, and Z coordinates in \AA .   
Orthogonal unit mesh translation vectors are 5.564 \AA\ in X and
5.176 \AA\ in Y.   }
\label{012hydrox}
\begin{tabular}{cccc}
atom&x&y&z\\
H1   &    1.763 &   0.441 &  -7.029  \\
H2   &    2.936 &   3.749 &   5.723  \\
H3   &    2.782 &   0.442 &   7.030  \\
H4   &    1.609 &   3.748 &  -5.723  \\
H5   &    0.000 &   0.068 &   7.030  \\
H6   &    4.391 &   1.937 &  -5.723  \\
H7   &    0.154 &   1.938 &   5.723  \\
H8   &    4.545 &   0.069 &  -7.029  \\
O1   &    1.070 &   0.429 &  -6.342  \\
O2   &    0.218 &   0.764 &   1.144  \\
O3   &    2.587 &   3.251 &   4.895  \\
O4   &    3.364 &   3.334 &  -2.612  \\
O5   &    2.663 &   0.510 &  -3.766  \\
O6   &    1.881 &   0.510 &   3.767  \\
O7   &    5.054 &   3.121 &   0.000  \\
O8   &    3.475 &   0.430 &   6.343  \\
O9   &    1.958 &   3.251 &  -4.895  \\
O10  &    1.181 &   3.334 &   2.612  \\
O11  &    4.327 &   0.764 &  -1.143  \\
O12  &    0.693 &   0.080 &   6.343  \\
O13  &    1.545 &   4.922 &  -1.143  \\
O14  &    4.740 &   2.434 &  -4.895  \\
O15  &    3.963 &   2.352 &   2.612  \\
O16  &    4.663 &   5.176 &   3.767  \\
O17  &    2.272 &   2.564 &   0.000  \\
O18  &    5.445 &   0.000 &  -3.767  \\
O19  &    0.582 &   2.352 &  -2.612  \\
O20  &    5.369 &   2.434 &   4.895  \\
O21  &    3.852 &   0.080 &  -6.342  \\
O22  &    3.000 &   4.922 &   1.143  \\
Fe1  &    0.749 &   1.546 &  -0.802  \\
Fe2  &    3.912 &   4.265 &  -4.733  \\
Fe3  &    3.149 &   4.113 &   2.971  \\
Fe4  &    1.396 &   4.113 &  -2.970  \\
Fe5  &    0.633 &   4.265 &   4.733  \\
Fe6  &    3.796 &   1.546 &   0.802  \\
Fe7  &    3.415 &   1.421 &   4.733  \\
Fe8  &    1.014 &   4.140 &   0.802  \\
Fe9  &    4.178 &   1.573 &  -2.970  \\
Fe10 &    1.130 &   1.421 &  -4.733  \\
Fe11 &    0.367 &   1.573 &   2.971  \\
Fe12 &    3.531 &   4.140 &  -0.802  \\
\end{tabular}
\end{table}
\begin{table}
\caption{Interplanar spacings (picometers) of
hydrated and hydroxylated hematite (012) 
(as shown in Figure~\protect\ref{012.hydrated}) compared 
with bulk and relaxed nonhydroxylated surface structures}
\label{012relaxhydrox}
\begin{tabular}{ccccc}
&bulk&relaxed&hydrated&hydroxylated\\
O &  &  &  & \\
  & 35 & 29 (-18\%) & 26 (-26\%) & 16 (-54 \%) \\
Fe&  &  &  & \\
  & 80 & 66 (-17\%) & 70 (-12\%) & 97 (21 \%) \\
O &  &  &  & \\
  & 80 & 84 (5\%)   & 82 (3\%)   & 80 (-1 \%) \\
Fe&  &  &  & \\
  & 35 & 46 (32\%)  & 45 (29\%)  & 36 (4 \%) \\
O &  &  &  & \\
  & 149 & 140 (-6\%)  & 141 (-5\%)  & 147 (-1 \%) \\
O &  &  &  & \\
\end{tabular}
\end{table}

\begin{table}
\caption{Interplanar spacings (picometers) of
hydrated and hydroxylated hematite (001)
(as shown in Figure~\protect\ref{001.hydrated}) compared 
with bulk and relaxed nonhydroxylated surface structures}
\label{001relaxhydrox}
\begin{tabular}{ccccc}
&bulk&relaxed&hydrated&hydroxylated\\
Fe &  &  &  & \\
  & 86 & 44 (-49\%) & 48 (-44\%) & 89 (4\%) \\
O&  &  &  & \\
  & 86 & 83 (-4\%) & 83 (-3\%) & 78 (-9\%) \\
Fe &  &  &  & \\
  & 63 & 39 (-37\%)   & 38 (-39\%)   & 42 (-32\%) \\
Fe&  &  &  & \\
  & 86 & 102 (19\%)  & 102 (19\%)  & 98 (14\%) \\
O &  &  &  & 
\end{tabular}
\end{table}
\begin{table}
\caption{Slab unit cell coordinates for the hydrated (001) surface.  
X, Y, and Z coordinates in \AA .   
Orthogonal unit mesh translation vectors are 8.976 \AA\  in X and
5.176 \AA\  in Y.  } 
\label{001hydrate}
\begin{tabular}{cccc}
atom&x&y&z\\
H1 &  4.989000  & 3.646000 & -6.607000 \\
H2 &  3.927000  & 4.686000 & -6.662000 \\
H3 &  0.400000  & 2.017000 & -6.664000 \\
H4 &  8.314000  & 0.966000 & -6.607000 \\
H5 &  6.296000  & 1.119000 &  6.663000 \\
H6 &  5.210000  & 2.132000 &  6.584000 \\
H7 &  1.875000  & 4.760000 &  6.608000 \\
H8 &  0.795000  & 3.739000 &  6.640000 \\
O1 &   0.000000 &  4.964000 & -1.193000 \\
O2 &  1.410000  & 2.409000 &  3.422000 \\
O3 &  4.490000  & 2.376000 & -1.190000 \\
O4 &  5.904000  & 4.997000 &  3.419000 \\
O5 &  7.366000  & 2.494000 &  3.416000 \\
O6 &  5.851000  & 0.000000 & -1.194000 \\
O7 &  1.364000  & 2.590000 & -1.189000 \\
O8 &  2.877000  & 5.086000 &  3.426000 \\
O9 &  8.800000  & 5.009000 &  3.407000 \\
O10&    7.231000 &  2.367000 & -1.197000 \\
O11&    2.743000 &  4.954000 & -1.186000 \\
O12&    4.314000 &  2.427000 &  3.423000 \\
O13&    4.316000 &  0.753000 & -3.429000 \\
O14&    5.731000 &  3.373000 &  1.179000 \\
O15&    8.810000 &  3.340000 & -3.434000 \\
O16&    1.245000 &  0.785000 &  1.181000 \\
O17&    8.857000 &  3.159000 &  1.177000 \\
O18&    7.344000 &  0.663000 & -3.435000 \\
O19&    2.855000 &  3.255000 & -3.428000 \\
O20&    4.369000 &  0.574000 &  1.183000 \\
O21&    5.906000 &  3.322000 & -3.435000 \\
O22&    7.477000 &  0.795000 &  1.174000 \\
O23&    1.420000 &  0.740000 & -3.421000 \\
O24&    2.990000 &  3.383000 &  1.185000 \\
O25&    4.447000 &  4.188000 & -5.980000 \\
O26&    8.867000 &  1.499000 & -5.979000 \\
O27&    5.779000 &  1.604000 &  5.969000 \\
O28&    1.324000 &  4.241000 &  5.969000 \\
Fe1&   4.359000 &  4.168000 & -3.910000 \\
Fe2&  5.859000  & 1.581000 &  0.293000 \\
Fe3&  8.851000  & 1.579000 & -3.909000 \\
Fe4&  1.372000  & 4.168000 &  0.295000 \\
Fe5&  5.856000  & 1.579000 & -2.601000 \\
Fe6&  7.356000  & 4.167000 &  2.199000 \\
Fe7&  1.368000  & 4.169000 & -2.596000 \\
Fe8&  2.867000  & 1.582000 &  2.205000 \\
Fe9&  8.848001  & 1.581000 & -0.307000 \\
Fe10 &  5.861000 &  1.582000 &  3.898000 \\
Fe11 &  1.369000 &  4.170000 &  3.898000 \\
Fe12 &  4.362000 &  4.168000 & -0.304000 \\
Fe13 &  8.853000 &  1.581000 &  2.584000 \\
Fe14 &  7.353000 &  4.167000 & -2.217000 \\
Fe15 &  2.865000 &  1.582000 & -2.211000 \\
Fe16 &  4.365000 &  4.171000 &  2.590000 \\ 
\end{tabular}
\end{table}
\begin{table}
\caption{Slab unit cell coordinates for the hydroxylated (001) surface.  
X, Y, and Z coordinates in \AA .   
Orthogonal unit mesh translation vectors are 8.976 \AA\ in X and
5.176 \AA\ in Y.   }
\label{001hydrox}
\begin{tabular}{cccc}
atom&x&y&z\\

H1   &  8.478001 &  5.121000 &  4.469000  \\
H2   &  3.991000 &  2.534000 &  4.470000  \\
H3   &  6.468000 &  3.444000 & -4.468000  \\
H4   &  1.980000 &  0.856000 & -4.469000  \\
H5   &  4.417000 &  4.322000 & -6.994000  \\
H6   &  8.902000 &  1.740000 & -6.994000  \\
H7   &  6.046000 &  1.651000 &  6.994000  \\
H8   &  1.547000 &  4.245000 &  6.994000  \\
O1   &  0.149000 &  5.068000 & -1.204000  \\
O2   &  1.585000 &  2.589000 &  3.353000  \\
O3   &  4.638000 &  2.480000 & -1.202000  \\
O4   &  6.073000 &  0.000000 &  3.355000  \\
O5   &  7.481000 &  2.621000 &  3.362000  \\
O6   &  5.992000 &  0.073000 & -1.164000  \\
O7   &  1.503000 &  2.660000 & -1.164000  \\
O8   &  2.993000 &  0.033000 &  3.363000  \\
O9   &  8.954000 &  5.145000 &  3.581000  \\
O10  &  7.321000 &  2.500000 & -1.156000  \\
O11  &  2.832000 &  5.086000 & -1.154000  \\
O12  &  4.467000 &  2.557000 &  3.582000  \\
O13  &  4.386000 &  0.800000 & -3.353000  \\
O14  &  5.819000 &  3.495000 &  1.203000  \\
O15  &  8.875000 &  3.390000 & -3.355000  \\
O16  &  1.331000 &  0.906000 &  1.202000  \\
O17  &  8.955000 &  3.317000 &  1.163000  \\
O18  &  7.465000 &  0.769000 & -3.362000  \\
O19  &  2.977000 &  3.356000 & -3.362000  \\
O20  &  4.467000 &  0.729000 &  1.165000  \\
O21  &  5.991000 &  3.418000 & -3.581000  \\
O22  &  7.626000 &  0.891000 &  1.155000  \\
O23  &  1.503000 &  0.830000 & -3.582000  \\
O24  &  3.137000 &  3.476000 &  1.157000  \\
O25  &  4.488000 &  4.316000 & -6.081000  \\
O26  &  0.000000 &  1.726000 & -6.081000  \\
O27  &  5.974000 &  1.661000 &  6.081000  \\
O28  &  1.482000 &  4.249000 &  6.080000  \\
Fe1  &  4.317000 &  4.316000 & -4.252000  \\
Fe2  &  5.988000 &  1.706000 &  0.358000  \\
Fe3  &  8.805000 &  1.729000 & -4.252000  \\
Fe4  &  1.500000 &  4.292000 &  0.357000  \\
Fe5  &  5.945000 &  1.611000 & -2.577000  \\
Fe6  &  7.426000 &  4.272000 &  2.153000  \\
Fe7  &  1.457000 &  4.197000 & -2.577000  \\
Fe8  &  2.938000 &  1.683000 &  2.155000  \\
Fe9  &  8.959000 &  1.685000 & -0.359000  \\
Fe10 &  6.142000 &  1.661000 &  4.252000  \\
Fe11 &  1.655000 &  4.249000 &  4.250000  \\
Fe12 &  4.471000 &  4.272000 & -0.357000  \\
Fe13 &  0.024000 &  1.779000 &  2.579000  \\
Fe14 &  7.520000 &  4.294000 & -2.152000  \\
Fe15 &  3.030000 &  1.703000 & -2.152000  \\
Fe16 &  4.512000 &  4.365000 &  2.579000  \\
\end{tabular}
\end{table} 
\end{document}